\documentclass[aps,prapplied,amsmath,amssymb,twocolumn,superscriptaddress]{revtex4-2}

\usepackage{graphicx}
\usepackage{svg}
\usepackage{adjustbox}
\usepackage{fixltx2e}
\usepackage{makecell}
\usepackage{hyperref}
\usepackage{cleveref}
\usepackage{subcaption}
\graphicspath{ {./pictures/} }

\begin{document}


\title{Wafer-Scale Characterization of Al/$\text{Al}_x\text{O}_y$/Al Josephson Junctions at Room Temperature}


\author{S. J. K. Lang}
\email[]{simon.lang@emft.fraunhofer.de}
\affiliation{Fraunhofer Institut für Elektronische Mikrosysteme und Festkörpertechnologien EMFT, Munich, Germany}
\author{I. Eisele}
\affiliation{Fraunhofer Institut für Elektronische Mikrosysteme und Festkörpertechnologien EMFT, Munich, Germany}
\affiliation{Center Integrated Sensor Systems (SENS), Universität der Bundeswehr München, Munich, Germany}
\author{J. Weber}
\affiliation{Fraunhofer Institut für Elektronische Mikrosysteme und Festkörpertechnologien EMFT, Munich, Germany}
\author{A. Schewski}
\affiliation{Fraunhofer Institut für Elektronische Mikrosysteme und Festkörpertechnologien EMFT, Munich, Germany}
\affiliation{Technical University of Munich, Munich, Germany}
\author{E. Music}
\affiliation{Fraunhofer Institut für Elektronische Mikrosysteme und Festkörpertechnologien EMFT, Munich, Germany}
\author{A. Maiwald}
\affiliation{Fraunhofer Institut für Elektronische Mikrosysteme und Festkörpertechnologien EMFT, Munich, Germany}
\author{M. Heigl}
\affiliation{Fraunhofer Institut für Elektronische Mikrosysteme und Festkörpertechnologien EMFT, Munich, Germany}
\author{D. Zahn}
\affiliation{Fraunhofer Institut für Elektronische Mikrosysteme und Festkörpertechnologien EMFT, Munich, Germany}
\author{L. Zhen}
\affiliation{Technical University of Munich, Munich, Germany}
\author{L. Nebrich}
\affiliation{Fraunhofer Institut für Elektronische Mikrosysteme und Festkörpertechnologien EMFT, Munich, Germany}
\author{B. Schoof}
\affiliation{Technical University of Munich, Munich, Germany}
\author{T. Mayer}
\affiliation{Fraunhofer Institut für Elektronische Mikrosysteme und Festkörpertechnologien EMFT, Munich, Germany}
\author{L. Sturm-Rogon}
\affiliation{Fraunhofer Institut für Elektronische Mikrosysteme und Festkörpertechnologien EMFT, Munich, Germany}
\author{W. Lerch}
\affiliation{Fraunhofer Institut für Elektronische Mikrosysteme und Festkörpertechnologien EMFT, Munich, Germany}
\author{R. N. Pereira}
\affiliation{Fraunhofer Institut für Elektronische Mikrosysteme und Festkörpertechnologien EMFT, Munich, Germany}
\author{C. Kutter}
\affiliation{Fraunhofer Institut für Elektronische Mikrosysteme und Festkörpertechnologien EMFT, Munich, Germany}
\affiliation{Center Integrated Sensor Systems (SENS), Universität der Bundeswehr München, Munich, Germany}




\date{\today}

\begin{abstract}
Josephson junctions (JJs) are the key element of many devices operating at cryogenic temperatures. Development of time-efficient wafer-scale JJ characterization for process optimization and control of JJ fabrication is essential. Such statistical characterization has to rely on room temperature techniques since cryogenic measurements typically used for JJs are too time consuming and unsuitable for wafer-scale characterization. In this work, we show that from room temperature capacitance and current-voltage measurements, with proper data analysis, we can independently obtain useful parameters of the JJs on wafer-scale, like oxide thickness, tunnel coefficient, and interfacial defect densities. Moreover, based on detailed analysis of current vs voltage characteristics, different charge transport mechanisms across the junctions can be distinguished. We exemplary demonstrate the worth of these methods by studying junctions fabricated on 200 mm wafers with an industrially scale-able concept based on subtractive processing using only CMOS compatible tools. From these studies, we find that our subtractive fabrication approach yields junctions with quite homogeneous “average” oxide thickness across the full wafers, with a spread of less then 3\%. The analysis also revealed a variation of the tunnel coefficient with oxide thickness, pointing to a stoichiometry gradient across the junctions’ oxide width. Moreover, we estimated relatively low interfacial defect densities in the range of $70-5000$ defects/cm$^2$ for our junctions and established that the density increased with decreasing oxide thickness, indicating that the wet etching process applied in the JJs fabrication for oxide thickness control leads to formation of interfacial trap states.
\end{abstract}


\maketitle
\section{Introduction}
Josephson junctions are the key element of many different devices, such as traveling wave parametric amplifiers, superconducting quantum interference devices and superconducting qubits, which operate at cryogenic temperatures. Characterization of JJs has been relying mostly on cryogenic measurements, which are very time consuming. Development of time-efficient and wafer-scale characterization of JJs is currently a major research topic. Such characterization should involve only room temperature techniques since cryogenic characterization typically applied to JJs is too time consuming and unsuitable for wafer-scale characterization. The main objective of this wafer-scale characterization is to predict the cryogenic behaviour of JJ devices, e.g. performance of superconducting qubits, from measurements carried out at room temperature, before the devices are cooled and measured in a cryostat. A time-efficient wafer-scale JJ characterization at room temperature will be particularly important for the implementation of JJ fabrication based on standard CMOS manufacturing tools \cite{VanDamme_Advanced_2024,lang_aluminum_2023}, which demands close process optimization and control at wafer scale. This CMOS-based fabrication of JJs is essential for scalability and enhanced reliability/precision towards industrialization and commercialization of JJ devices. A recent study \cite{VanDamme_Advanced_2024} showed that with an industry-compatible fabrication of Al/$\text{Al}_x\text{O}_y$/Al JJs, transmon-style qubits with a high yield of 98.5 \% and $T_1$ times of up to 167 µs could be fabricated, approaching values typically achieved with conventional laboratory-style fabrication based on double angle evaporation techniques \cite{place_new_2021, biznarova_mitigation_2024, ganjam_surpassing_2023}. In the study, the authors used room temperature wafer-scale measurements to study their JJs, from which they could unveil for example that the variability of their qubit devices in terms of frequency, which is limited by the current through the JJs, is determined by variations in the JJs oxide barrier height instead of the junction area. The study suggested that oxide thickness and barrier height are critical parameters, but which of the two dominates and how their variability can be characterized and controlled during fabrication remains open. A deeper understanding of such questions requires thorough wafer-scale statistical analysis of JJs, with JJ parameters varied in a well-defined manner. 

In the present study, we show that useful JJ parameters (e.g. oxide thickness, effective barrier height, and interfacial defect densities) can be obtained on wafer-scale from simple room temperature capacitance and current-voltage measurements and proper data analysis. Based on detailed analysis of current vs voltage characteristics, different charge transport mechanisms across the junctions can also be distinguished. These analytical methods are exemplary applied to study junctions fabricated on 200 mm wafers with an industrially scalable subtractive processing based on CMOS tools similar to that reported in recent studies \cite{VanDamme_Advanced_2024,lang_aluminum_2023}. From the analysis, we could gain insight for example on oxide thickness values and wafer-scale homogeneity, on oxide barrier height and related stoichiometry gradients, as well as on interfacial defect densities. Our study exemplifies the usefulness of the proposed characterization principles for studying JJs and their wafer-scale parameter distributions, which is essential for time-efficient process control and optimization at industrial environment.

\section{Experimental}
\begin{figure}[h!]
   \centering
    \includegraphics[width=1.2\linewidth]{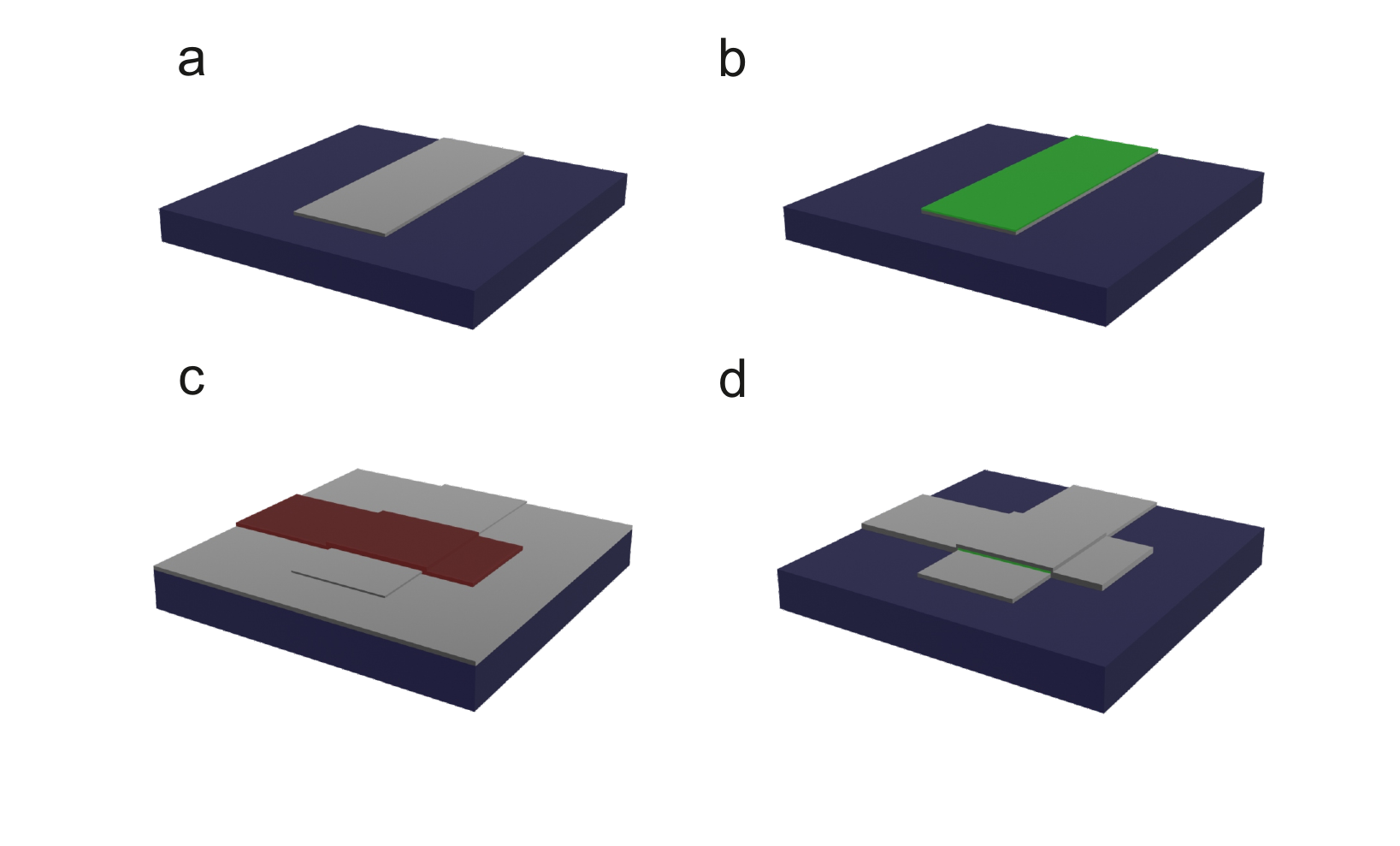}
    \caption{Scheme of the fabrication process of the JJs in the CMOS pilot line: a. Bottom Al (grey) electrode after structuring on silicon (blue), b. Wet chemical oxide etching (green), c. Top Al deposition and lithography of second layer, d. Structuring of top electrode.}
    \label{fig:Schematic}
\end{figure}

The fabrication process of our CMOS-compatible Al/$\text{Al}_x\text{O}_y$/Al JJs is quite similar to a previously reported process \cite{lang_aluminum_2023}. However, a few variations where implemented, as described below, to enable a wider range of oxide thicknesses, required for the present study. The substrates consisted of high ohmic (100) p-type 200 mm silicon wafers with a resistivity 3-5 k$\Omega$cm. Prior to deposition of the first Al layer, an ex-situ liquid HF dip for 3 min is performed for removal of the native silicon oxide. In a cluster-tool (Clusterline 200), after an initial pre-heating step, 120 nm of Al is deposited. The bottom electrode layer is lithographically patterned using an i-line stepper (Canon). A dry etch step is performed in an AMAT P5000 system using a chlorine-based etch gas. We note that during the photoresist strip, which is conducted using H\textsubscript{2}O plasma in the same device as the dry etching, the surface of the bottom Al electrode becomes further oxidized beyond the native aluminum oxide that results from the vacuum breakage during lithography (Fig. \ref{fig:Schematic}a). The thickness of the Al surface oxide resulting from the H\textsubscript{2}O plasma has been measured with ellipsometry to be around 4.4 nm. Ellipsometry is an optical method used to study the dielectric properties of thin films by analyzing the change in the polarization of light due to reflection or transmission, and comparing these changes to a theoretical model. This technique has become the standard for determining the thickness of silicon oxide layers in the semiconductor industry.

We actually use this oxide as a reference $\text{Al}_x\text{O}_y$ layer in the JJs. To achieve oxides with different thicknesses, we performed subsequent wet-etching in phosphoric acid for different times (10 s, 20 s, 30 s) (Fig. \ref{fig:Schematic}b) to manufacture our samples (etch10, etch20 and etch30; see Tab. 1). For the reference sample, no wet etching of the H\textsubscript{2}O plasma oxide was performed. The etching process is followed by the deposition of the top electrode. It consists of 100 nm Al deposited in the cluster-tool similar to the bottom layer to complete the JJ (Fig. \ref{fig:Schematic}c). During the etching of the top layer (Fig. \ref{fig:Schematic}d), precautions are taken to avoid etching through the bottom Al electrode as the process lacks selectivity.

\begin{figure}[h!]
\centering
    \begin{subfigure}[b]{0.95\linewidth}
        \centering
        \includegraphics[width=\linewidth]{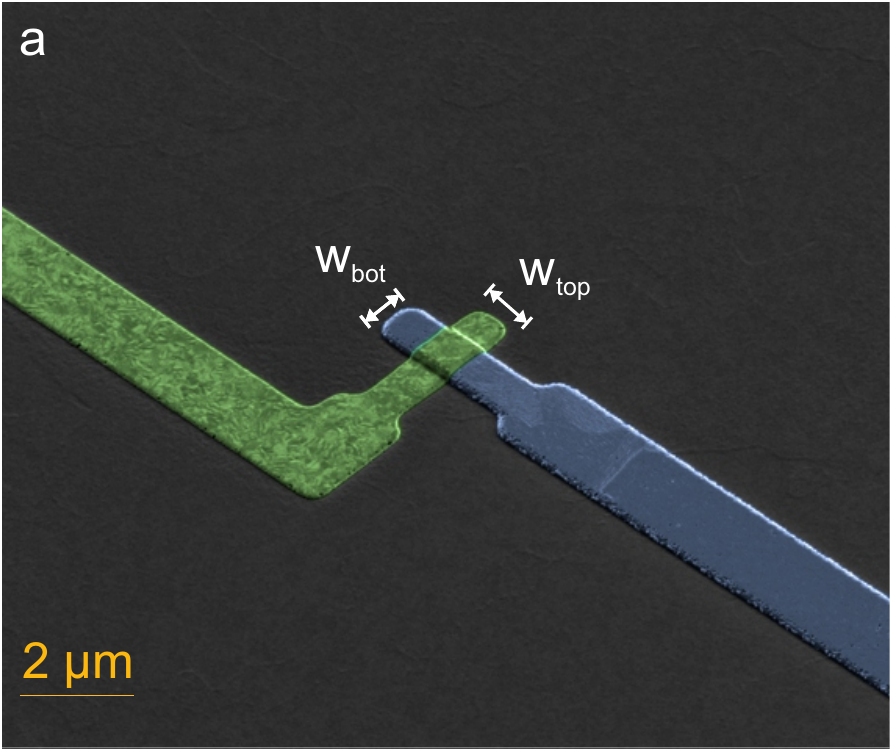}
        \label{fig:1sub1}
    \end{subfigure}
\vskip\baselineskip
    \begin{subfigure}[b]{0.95\linewidth}
        \centering
        \includegraphics[width=\linewidth]{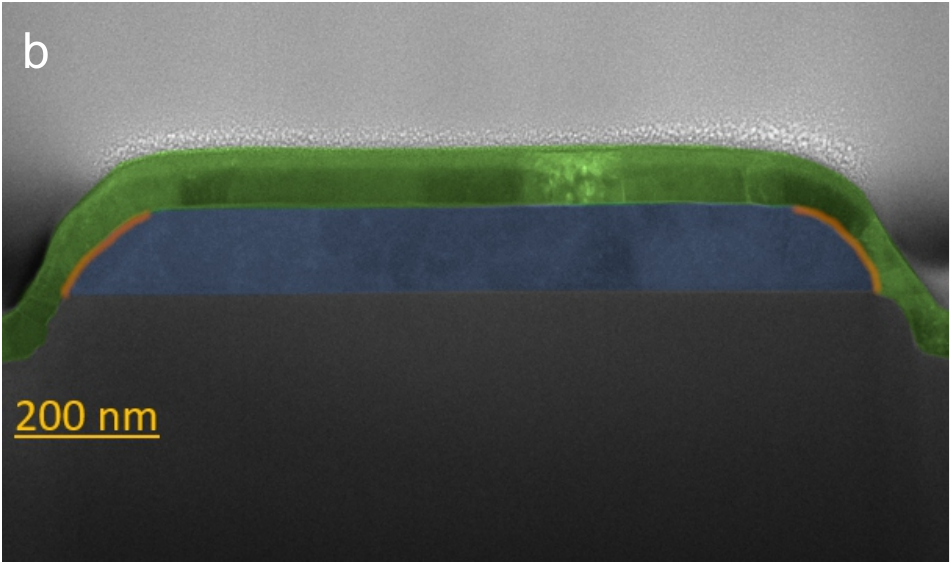}
        \label{fig:1sub2}
    \end{subfigure}
\caption{a. Top view of a 1 µm $\times$ 1 µm JJ with top (green) and bottom (blue) electrodes. b. TEM cross section of the same JJ with orange-marked sidewall region of the junction overlap area}
\label{fig:SEM}
\end{figure}

To analyze the leakage current of the JJs, test structures with different sizes, e.g. varying widths of the top ($\textnormal{w}_{top}$) and bottom ($\textnormal{w}_{bot}$) electrodes were designed (Fig. \ref{fig:SEM}a). The resulting overlap area of a JJ can be simplified to the top area, defined as $A = \textnormal{w}_{top} \cdot \textnormal{w}_{top}$. Later on, the sidewall area (orange region, Fig. \ref{fig:SEM}b), defined as $A_S = 2 \cdot h \cdot \textnormal{w}_{top} $, where $h$ is the thickness of the bottom Al layers, will be taken into account for the detailed analysis of the current flow across the junction. Comparing the current flow for different samples with varying geometries enables a comprehensive oxide analysis. In total, 256 different test structures with $A$ ranging from 0.12 µm\textsuperscript{2} to 1600 µm\textsuperscript{2} were included in each single chip. 

The electrical measurements were conducted in a fully automated wafer prober, which could handle an entire box of wafers. The layout for the stepping could be directly extracted from the lithography process. It is important to emphasize that without calibration test structures that included all the metal lines except for the junction crossing, it would have been impossible to achieve high measurement accuracy in the capacitance measurements down to $fF$. These structures are essential for enabling the effective assessment of even very small junction areas ($\sim$ 5 µm$^2$), which would otherwise remain undetectable. 

For the reference wafer, thin lamella ($\sim$ 50 nm) were extracted from the finalized JJs located at the wafer center for the structural analysis of the oxide thickness $t^{EELS}_{ox}$ by the oxygen concentration via TEM (transmission electron microscopy) and EELS (electron energy loss spectroscopy) measurements in a SEM (scanning electron microscope). The EELS technique measures the energy loss experienced by electrons as they interact with atoms while traversing a thin lamella. In our case, this method enables the identification of the oxygen position and content in the JJs. TEM cross-section images are formed by the interaction of the electron beam with the transmitted lamella. As high energy electrons have a minor wavelength compared to optical light, even features down to nm sizes are identifiable with this method.

\section{Results and Discussion}

\subsection{Wafer-scale Capacitance Measurements}
\begin{figure}[h!]
    \centering
    \begin{subfigure}[b]{0.91\linewidth}
        \centering
        \includegraphics[width=\linewidth]{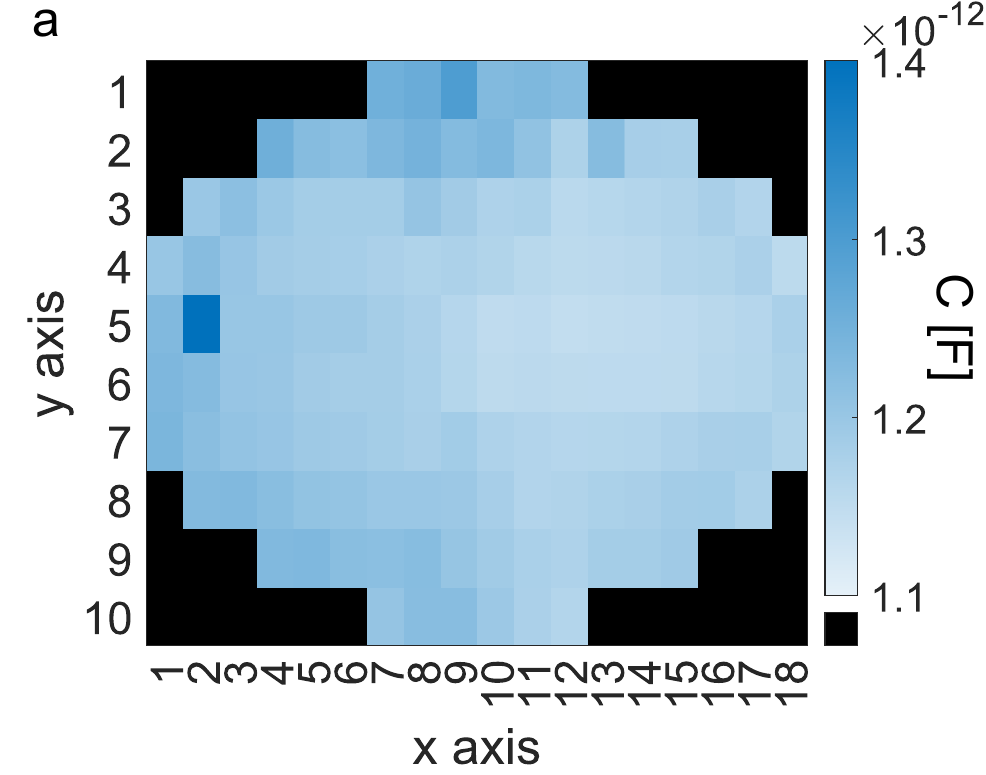}
        \label{fig:5sub1}
    \end{subfigure}
    \vskip\baselineskip
    \begin{subfigure}[b]{0.87\linewidth}
        \centering
        \includegraphics[width=\linewidth]{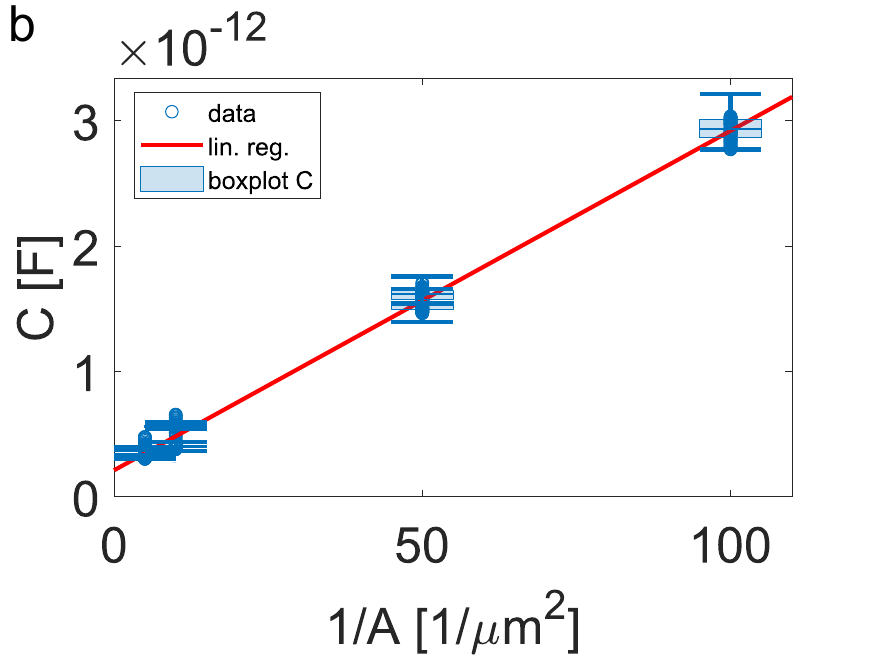}
        \label{fig:5sub2}
    \end{subfigure}
    \vskip\baselineskip
    \begin{subfigure}[b]{0.84\linewidth}
        \centering
        \includegraphics[width=\linewidth]{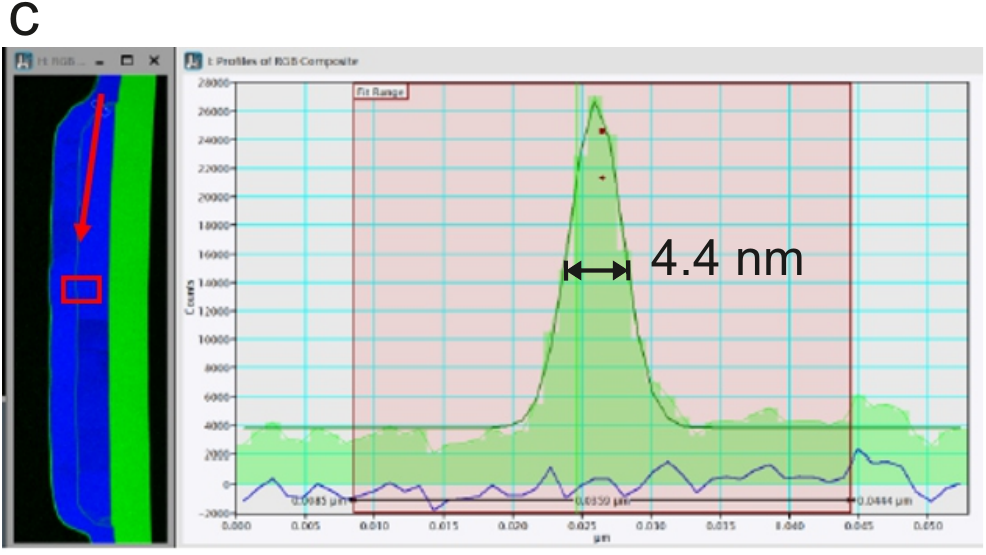}
        \label{fig:5sub3}
    \end{subfigure}
\caption{a. Across-wafer capacitance map with 140 junctions of area $A$ = 50 µm\textsuperscript{2} with $\bar{C} = 1.19$ pF, relative standard deviation $RSD_{\bar{C}} = 2.3$ \% and functional devices of $98.6$ \%. b. Capacitance ${C}$ vs. area $A$ for the reference wafer with a linear regression (the capacitances of the single JJs are indicated with circles). c. Oxygen distribution measured with EELS of a junction for the reference sample. Measurement spot indicated with arrow (red) between the top and bottom aluminum electrodes (blue). FWHM of the gaussian oxygen peak was measured to be $t^{EELS}_{ox} = 4.4 $ nm }
\label{fig:CM}
\end{figure}

 Wafer-scale capacitance measurements were conducted for junctions with varying overlap areas $A$ between 1 µm\textsuperscript{2} and 1600 µm\textsuperscript{2}. In detail, a mean capacitance $\bar{C}$ was calculated from across-wafer capacitance measurements of a specific A. The relative standard deviation (RSD) of the capacitance was below  $RSD_{\bar{C}} < 3$ \% for all samples. An example of a capacitance map is shown in Fig. \ref{fig:CM}a. By using multiple overlap areas and applying a simple plate capacitor model in Eq. (1), the ratio ${C}/A$ can be obtained analogue for all samples (see Tab. 1) by a linear regression with a crossing with the y-axis close to the origin, i.e. design and fabricated areas agree with a minor offset. An example is shown for the etch20 sample in Fig. \ref{fig:CM}b.

\begin{equation}
{C}/A = \frac{\varepsilon_r \varepsilon_0}{t_{ox}}
\end{equation}
The dielectric constant $\varepsilon_r$ can be calculated directly from the ${C}/A$, with $\varepsilon_0$ being the vacuum permittivity, if the oxide thickness $t_{ox}$ is known. Therefore, EELS measurements on a single junction of the reference wafer's center were performed. By analysing the FWHM of the oxygen Gaussian distribution, $t^{EELS}_{ox} = 4.4 $ nm (Fig. \ref{fig:CM}c) was calculated, which is in agreement with ellipsometry measurements (see experimental section). For the reference JJs, this resulted in  $\varepsilon_r \sim 10 $, which is higher than the commonly known value of $\varepsilon_r$ $\sim$ 8, but still in agreement with other investigations \cite{kolodzey_electrical_2000}. Assuming that $\varepsilon_r$ remains approximately constant for all samples measured in this work, $t_{ox}$ for the etched samples could be calculated from the capacitance measurements using Eq. (1) (see Tab. 1). In this manner, $\bar{t}_{ox}$ is the across-wafer mean value with a relative standard deviation across the wafer of $RSD_{\bar{t}_{ox}} \sim  RSD_{\bar{C}}  < 3$ \% defined by the capacitance measurement. This makes the capacitance measurements a highly accurate, time efficient process monitoring technique during wafer processing, not only for single junction, but also for the entire wafer, compared to the known time-intensive single die TEM measurements \cite{zeng_direct_2015}.

\subsection{Current-Voltage (I-V) Characteristics}

In recent years, room temperature (RT) measurements have become a common method to determine the approximate DC resistances ($\sim$ 10 k$\Omega$) of Josephson junctions \cite{Osman_2021, pishchimova_improving_2022}. Using the Ambegaokar-Baratoff relation \cite{ambegaokar_tunneling_1963}, one can directly infer the critical current of the junction, which represents its Josephson energy. Since this value, along with the capacitive energy, determines the qubit frequency ($\sim$ 4.5 GHz) and the noise dependency of the qubit, it is a central parameter for the qubit characterization, in particular its frequency targeting.

\begin{figure}[h!]
   \centering
   \includegraphics[width=0.9\linewidth]{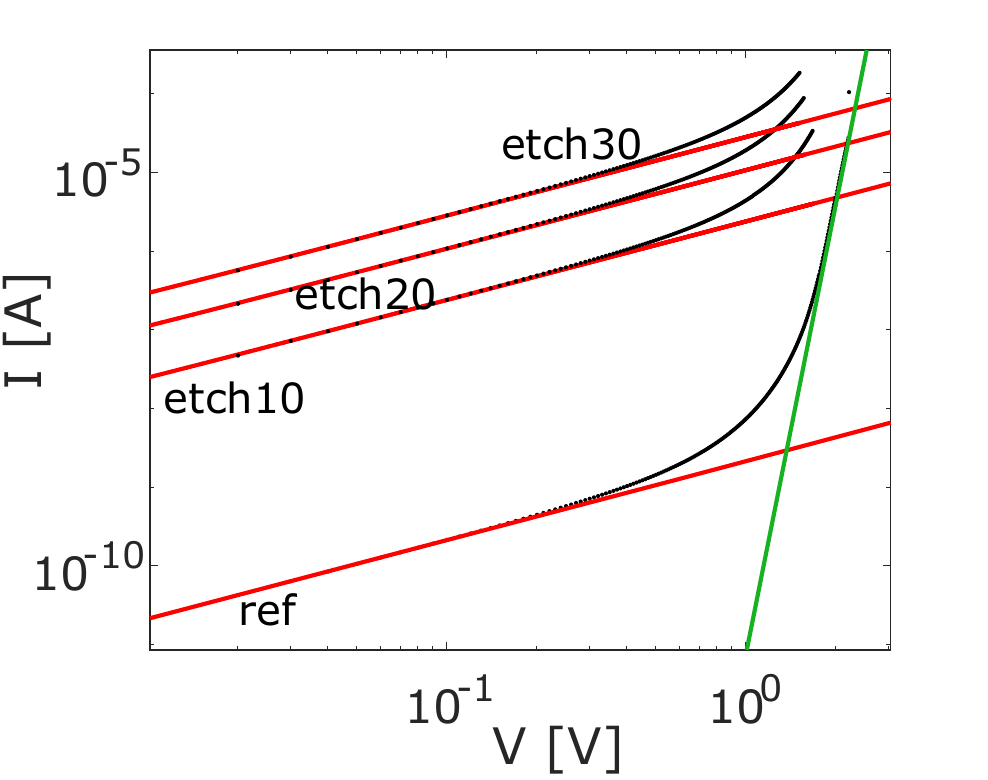}
    \caption{I-V curves of JJs with size $ A = $25 µ m$^2$ from the different wafers center with fitted DT (red) and FN (green)}
    \label{fig:IV}
\end{figure}

In our study, current-voltage IV curves have been recorded to investigate the JJs tunnel transport in detail. The IV characteristics can be obtained by gradually increasing the voltage across the junction while measuring the corresponding current flow. Slow sweeping rates are important for high accuracy. As examples, the room temperature electrical data of the reference and the etched JJs are shown in Fig. \ref{fig:IV} (measurements taken from JJs in the center of the wafers). 

An examination of the I-V curves indicates the presence of multiple tunneling mechanisms, making it essential to account for different tunneling regimes. At low voltage $V$ ($<$ 0.3 V), direct tunnelling (DT) appears for all samples (Fig. \ref{fig:IV}, red line) as it is the case for trap free insulators \cite{simmons_generalized_1963}:
\begin{equation}
I_{DT} =\frac{\alpha k AV}{ t_{ox}} \exp(-k t_{ox})
\end{equation}
with \(\alpha = \frac{e}{8\beta^2\pi^2 \hbar}\), tunnel coefficient \( k = 2\beta\sqrt{\frac{2\Phi m'}{\hbar^2}} \), average barrier height $\Phi$ of the oxide, effective electron mass $m’$, electron charge $e$ and $\beta$, a correction factor for a  non-rectangle shape of the potential barrier. With the wafer-mean oxide thickness $\bar{t}_{ox}$, $k$ can be fitted for the different samples (Tab. 1). As $k$ is depending on both, $\beta$ and $\phi$, we can only estimate $\phi$ for the reference sample. Here, EELS measurements (Fig. \ref{fig:CM}c) revealed a gaussian-shaped oxygen distribution across the junction oxide barrier. Therefore, we assume a comparable gassian-shape of the potential barrier, which we calculate to $\beta \sim $ 1. Therefore, $\phi \sim $ 3.14 eV for the reference sample using Eq. (2) with an effective electron mass of $m'$ = 0.75 $m_e$. The value is in agreement with literature \cite{cornette_relation_2020,gloos_properties_2003}. 

This approach, however, only applies to gaussian-shaped potential barriers, which may no longer be applicable to the etched samples. During the H\textsubscript{2}O plasma strip, we first perform a plasma densification of the oxide, which is modified during the wet-etch afterwards. As the oxide begins to etch at the surface, the less densified (i.e. less stoichiomertic) region is etched first, resulting in an overall increase in $k$ (see Tab. 1).
Given the non-homogeneous shape of the barrier, we should expect that both $\phi$ and $\beta$ change simultaneously, making it impossible to determine the individual impact of etching on each parameter.

Through further investigating the I-V curves, additional insights into traps in the oxide can be obtained. If traps are absent in oxides, at higher voltage electrons will accumulate at the metal insulator interface as space charges and create an internal electric field. Thus, the current flow will be enhanced according to Mott-Gurney’s law \cite{spahr_regimes_2013}  
\begin{equation}
I_{MG} = \frac{9}{8} \frac{A \mu \varepsilon_0\varepsilon_r V^2 }{t_{\text{ox}}^3}
\end{equation}
with the electron mobility $\mu$. However, in real materials the presence of traps cannot be neglected. The so-called space charge limited current (SCLC) is defined by the initial filling of shallow traps with electrons (shallow trap-squared law), which can be described by multiplying a trap factor $\theta$ to Eq. (3). Additional transport channels like (multi step) trap assisted tunneling (TAT) and Poole-Frenkel emission (PFE) will appear in parallel. PFE arrives from trapped electrons in the insulator, which get thermally released by the conduction band bending caused by an applied voltage \cite{spahr_conduction_2013}. Additionally, traps in the oxide facilitate tunneling of electrons by reducing the tunnel distance. As single trap assisted tunneling events are improbable, a collective approach is commonly applied to take multiple intermediate tunneling processes into account. The modern SCLC (mSCLC), combining SCLC with PFE and TAT, describes the electron transport in oxides in a good approximation with a simple power-law
\begin{equation}
I_{mSCLC} \sim V^m
\end{equation}
where, a power $m > 2$ implies the deviation from the idealised Mott-Gurney's law. For all our samples, this behaviour cannot be observed (Fig. \ref{fig:IV}), indicating the absence of traps in the oxide, prior to transitioning into the Fowler-Nordheim (FN) tunneling. This phenomenon is attributed to the reduction of the effective tunnel oxide width for electrons with the band bending of the AlOx conduction band beneath the metal band edge, thereby enhancing current flow \cite{gericke_fowler-nordheim_1970}
\begin{equation}
I_{FN} \sim \frac{V^2}{\Phi t_{ox}^2} \exp\left(-\frac{b t_{ox} \Phi^{3/2}}{V}\right)
\end{equation}                            
with the FN coefficient $b$ = 6.83 eV$^{-3/2}$ V/nm \cite{liang_quantum_2014}. At elevated voltages ($\sim$ 1 V), all samples exhibit the same gradient, which can be attributed to this tunneling process. In Fig 4., the curve fit (green) was tried, which resulted in $\Phi$ = 1.04 eV, which is a factor of 3 off from literature values \cite{cornette_relation_2020,gloos_properties_2003}. We conclude, that the premature breakthrough prevents the accurate fitting of $\Phi$ from FN tunneling, as typically done.

\subsection{Leakage Current Measurements}

Resistance measurements were carried out on JJs with different overlap sizes across the central region of the wafers. For a more in depth analysis, the simplified junction area $A = \textnormal{w}_{top}$ $\cdot$ $\textnormal{w}_{bot}$ (top area) is now extended with the sidewall area $A_S = 2 \cdot h \cdot \textnormal{w}_{top}$, whereby $h$ defines the height of the bottom Al layer, $\textnormal{w}_{top}$ the width of the top electrode and $\textnormal{w}_{bot}$ the width of the bottom electrode (see experimental section). This allows the analysis of the location of leakage (sidewall) current. For the assumption of cross-wafer homogeneous oxide thickness and resistivity \cite{sze_semiconductor_2012}, the top area-resistance ($RA$) and the sidewall area-resistance ($RA_{S}$) for the junctions can be defined for the two regions, separately. For the junction resistance, this results in
\begin{equation}
\frac{1}{R} = \frac{\textnormal{w}_{top} \cdot \textnormal{w}_{bot}}{RA} + \frac{2 \cdot h \cdot \textnormal{w}_{top}}{RA_{S}}     
\end{equation}
which transforms into
\begin{equation}
{R} = \frac{RA \cdot RA_{S}}{ \textnormal{w}_{bot} \cdot RA_{S} + 2 \cdot h \cdot RA }  \cdot\frac{1}{ \textnormal{w}_{top}}          
\end{equation}

For the condition of $\textnormal{w}_{bot} \cdot RA_{S}$ \(>>\) $2\cdot h \cdot RA$, which is satisfied when $\textnormal{w}_{bot}$ is sufficiently large, the equation can be further simplified as follows
\begin{equation}
R = \frac{RA}{\textnormal{w}_{top}\cdot \textnormal{w}_{bot}}
\end{equation}

\begin{figure}
   \centering
    \includegraphics[width=0.9\linewidth]{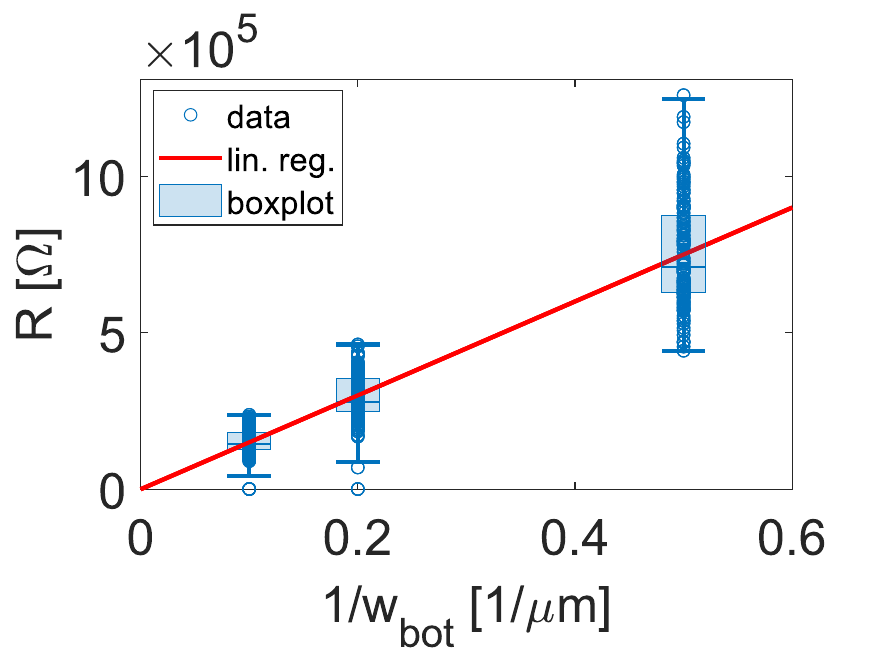}
    \caption{Cross-wafer JJ resistances potted against their different bottom junction width 1/$\textnormal{w}_{bot}$. As an example, data of wafer $etch30$ is shown}
\label{fig:LC}
\end{figure}

For the analysis of the region contribution to the resistance, Eq. (8) was used to fit $RA$ while holding $\textnormal{w}_{top}$ constant (see Fig. \ref{fig:LC}, red line). This ensures that the condition for Eq. (8) is fulfilled, as $1/\textnormal{w}_{bot}$ is further increased. Analogue to that, $RA_{S}$ is calculated by using $RA$ and fitting the pre-factor in Eq. (7), while holding $\textnormal{w}_{bot}$ constant. The results of $RA$ and $RA_{S}$ (see Tab. 1) are quite similar for the same samples, indicating the absence of sidewall effects and a homogeneous flow of the current across the top and sidewall area. By comparing the different samples with each other, $RA$ and $RA_{S}$ reduced with decreasing the oxide thickness by prolonged oxide etching while maintaining their comparability (Tab. 1). This behaviour defines our etching as a homogeneous oxide reduction process.

\begin{table*}[t]
	\begin{adjustbox}{width=\textwidth,center}
		\begin{tabular}{|c|c|c|c|c|c|c|c|c|c|c|c|}
			\hline
			\makecell{ \hfill\\ sample\\ \hfill }  	
			&\makecell{etch \\{[}s{]} }
    	    &\makecell{${C}/A$ \\ {[}fF/µm$^2${]}}
                &\makecell{\(\bar{t}_{ox}\)   \\{[}nm{]}}
			&\makecell{ $k$ \\{[}nm$^{-1}${]} }
			& \makecell{$RA$   \\{[}M$\Omega$µm$^2${]}}     
			& \makecell{$RA_{S}$ \\{[}M$\Omega$µm$^2${]}}
			&\makecell{$\bar{V}_{BT}$ \\{[}V{]}}

			&\makecell{ $E_{crit}$ \\{[}MV/cm{]} }
			&\makecell{ $D_{crit}$ \\{[}defects/cm$^2${]} }
			\\ \hline
			reference    & -    & 20.07 $\pm$ 0.03 & 4.4  & 15.7   & (10.9 $\pm$ 0.6) $\cdot10^3$ & (8.9 $\pm$ 0.8) $\cdot10^3$    & 2.18 $\pm$ 0.06  & 4.5         & 70 \\ \hline
			etch10& 10 & 25.99  $\pm$ 0.06 & 3.5   & 17.8 & 13.5 $\pm$ 0.6 & 5.9 $\pm$ 0.1 & 1.66 $\pm$ 0.07   & 4.4          & 1700   \\ \hline
			etch20 & 20 & 27.08 $\pm$ 0.06 & 3.3    &18.4  & 2.9 $\pm$ 0.1 & 2.17 $\pm$ 0.04  & 1.56 $\pm$ 0.07   & 4.3   & 2100  \\ \hline
			etch30 & 30  & 28.9 $\pm$ 0.1 & 3.1 &19.3 & 1.50 $\pm$ 0.03   & 1.24 $\pm$ 0.02  & 1.51 $\pm$ 0.07  & 4.6    & 5500  \\ \hline
		\end{tabular}
	\end{adjustbox}
	\caption{Parameter values derived for all samples from the room temperature electrical characterization data.}
\end{table*}

\subsection{Wafer-scale Breakthrough Measurements}

\begin{figure}[h!]
    \centering
    \begin{subfigure}[b]{0.93\linewidth}
        \centering
        \includegraphics[width=\linewidth]{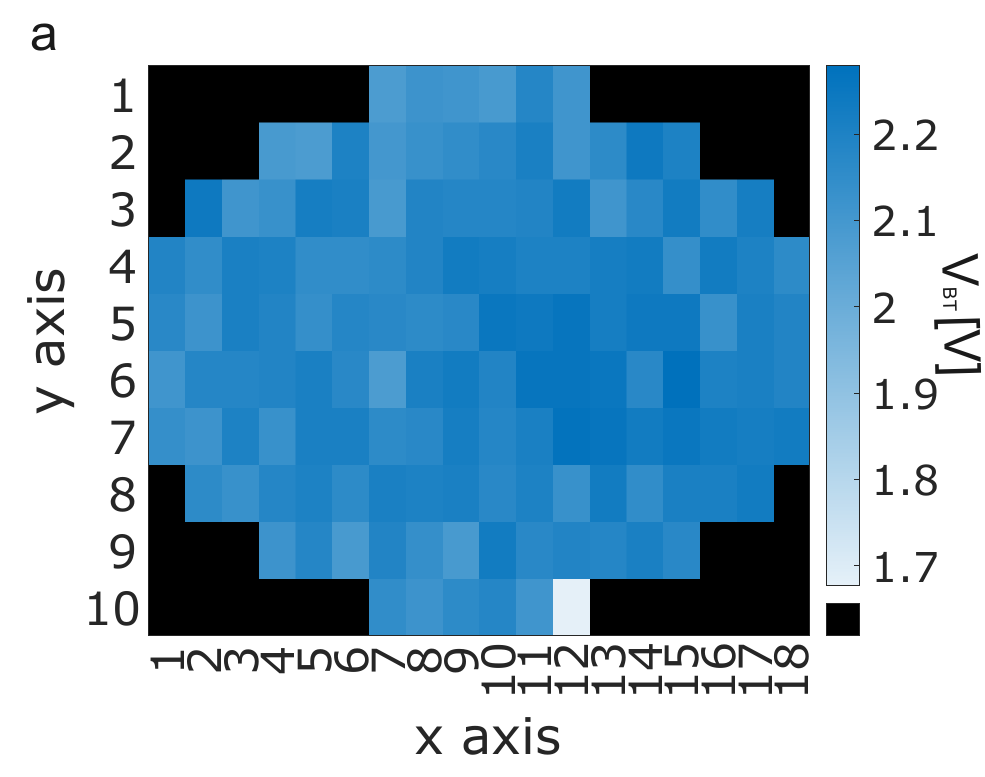}
        \label{fig:7sub1}
    \end{subfigure}
    \vskip\baselineskip
    \begin{subfigure}[b]{0.9\linewidth}
        \centering
        \includegraphics[width=\linewidth]{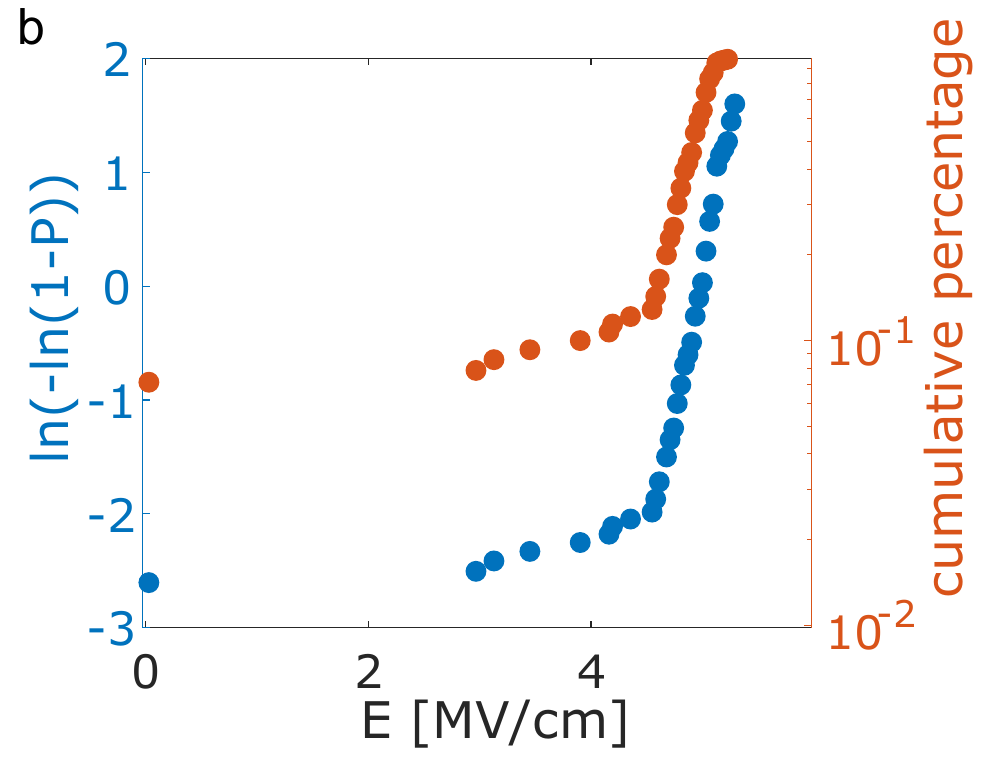}
    \end{subfigure}
\caption{a. Cross-wafer measurements of the breakthrough voltage for the reference sample. b.\(ln\{-ln(1 - P)\}\) (blue) and $P$ (red) vs. $E$ for the reference sample. At $E_{crit}$ = 4.6 MV/cm, the transition appears for $P_{k}$ = 12.9 \%.}
\label{fig:BT}
\end{figure}

In addition to resistance and capacitance measurements, the behavior of the junction can also be analyzed by observing its breakthrough voltage $V_{BT}$. For the measurements, the voltage was ramped up at a rate of 0.07 V/s in 10 mV increments. Wafer-scale measurements were performed on junction sizes of 5$\times$5 µm\textsuperscript{2} (see Fig. \ref{fig:BT}a). The measured across-wafer mean breakthrough voltages $\bar{V}_{BT}$ drops already from 2.18 V to 1.66 V (Tab. 1) by applying 10 sec oxide etch. For additional etching time, only a minor decrease in $\bar{V}_{BT}$ is recognisable. The relative standard deviation of the breakthrough mean was $ RSD_{\bar{V}_{BT}}\leq 4.6$ \% for sample. We note that this can only be measured accurately with a small stepsize during the ramping of the applied voltage.

The breakthrough voltage provide even more valuable insights into oxide behaviour by determination of the defect density at the junction interface. If thickness inhomogeneties occur or defects are located at the oxide interface, electrical breakthrough can appear at far lower voltages (defect related) than without having weak links (intrinsic). For standardized industrial processes, only a few defects occur, and for a statistical analysis, a large number of junctions need to be analysed to distinguish between defect-related and intrinsic breakthrough. This analysis employs a probability calculation similar to that used for the breaking of chains at the weakest link. A junction can be seen again as a simple plate capacitor and  subdivided in parallel paths; the overall capacitor will experience failure when the weakest capacitor in the grid breaks through. Based on the analysis of MOS time depended oxide breakthrough, the cumulative number of failed capacitors $P$ up to a certain voltage serves as an indicator for the transition from defect-related to intrinsic breakthrough \cite{verweij_dielectric_1996}. 

Those studies focus on the analysis of SiO\textsubscript{2} as an oxide barrier for MOS devices, whereas our approach is accessing Al\textsubscript{x}O\textsubscript{y} for JJ fabrication. At the transition point of the two regimes, the critical defect density $D_{crit}$ leading to the defect related breakthrough can be derived as well as the corresponding critical electric field-strength $E_{crit}$. For the analysis, the electrical field-strength $E = V_{BT} /\bar{t}_{ox} $ can be calculated from the applied breakthrough voltage and the wafer-mean oxide thickness. However, the point of transition, i.e $E_{crit}$ is reached, can only be seen by plotting $E$ in a Weibull graph, see Fig. \ref{fig:BT}b. We find for all our data a comparable $E_{crit} = 4.4 $ MV/m$^2$, which is in accordance with literature \cite{kolodzey_electrical_2000} and further confirming the accuracy of this approach. For the assumption of random defect spread, a Poisson distribution can be used. The critical defect density $D_{crit}$ results from the probability of finding defect-free capacitors
\begin{equation}
D_{crit} = \frac{-ln(1-P_k)}{A}                              
\end{equation}
with $P_{k}$ the probability at the transition \cite{wolters_behaviour_1987}. Our samples showed for the reference JJs a critical defect density of 70 cm\textsuperscript{-2}, which increased for decreasing oxide thicknesses by etching (Tab. 1). The increase in $D_{crit}$ could be attributed to an increase of the oxide surface roughness with increasing etching time. However, this would imply an increase of the roughness by a factor of 80, which is unrealistic. Therefore, we conclude that the increase in $D_{crit}$ should be due to an increased amount of surface traps induced by the wet-etching. Nonetheless, we note that the trap densities estimated for our junctions are in all cases relatively low (up to only 5500 defects/cm$^2$).

Defects created during etching, particularly in the JJs, have a negative impact on the homogeneity of the interface and render the structure unsuitable for qubit fabrication. Our work shows that process optimization and quality parameters relevant for quantum application at cryogenic temperatures can be discerned using electrical measurements at room temperature. Defects and other perturbations can cause increased losses and are possible sources of two-level states (TLS) \cite{de_graaf_two-level_2020}.

\section{Conclusion}
Using various test structures, $\text{Al}_x\text{O}_y$ JJs have been characterized at room temperature. Wafer-scale capacitance measurements, combined with electron energy loss spectroscopy, enabled a reliable determination of the effective oxide thicknesses of junctions across full 200 mm wafers. From this analysis, we find that our subtractive fabrication approach yields junctions with quite homogeneous “average” oxide thickness across the full wafers, with a thickness spread of less than 3\%, which to our knowledge has been measured for the first time. This result is particularly important for achieving highly precisie qubit frequency targeting on full 200 mm wafers. 

Moreover, analysis of leakage currents, by taking into account different contributions to the current across the whole junction interface, enabled us to infer about oxide uniformity within each junction. From this, we confirm a very uniform control of oxide thickness attained with our wet-etching approach. Based on detailed analysis of current-voltage characteristics, different charge transport mechanisms across the junctions could be distinguished. At low voltages the transport is ohmic, associated to direct tunnelling across the oxide barrier. At higher voltages, the onset of Fowler-Nordheim tunnelling is recognizable. As no intermediate regime is identified, trap-assisted tunneling processes are absent, i.e. high quality oxides can be prepared by the combination of plasma processing and wet-chemical etching. From the direct tunneling regime, we could calculate the tunnel coefficient $k$ for all oxide thicknesses. We found that $k$ increases with decreasing oxide thickness, pointing to a stoichiometry gradient across the oxide barrier width. Additionally, we could show that detailed statistical analysis of breakthrough voltages of JJs can provide valuable wafer-level information about interface defect densities. For our junctions, we calculated very low defect densities in the range  $70 - 5500$ defects/cm$^2$. The density increases with decreasing oxide thickness (increasing etching time), indicating that wet etching is involved in the formation of defects. Our study exemplifies the worth of the proposed room temperature characterization techniques for obtaining parameters of JJs and their distribution across full 200 mm wafers, which are useful for time-efficient process control and optimization at industrial environment.

\section*{Acknowledgements}
The authors would like to thank Chawki Dhieb for his support in electrical measurement automatization. We also appreciate Luca Rommeis for his support in text formatting and the whole Fraunhofer EMFT clean room staff for the professional fabrication. We acknowledge helpful discussions with G. Huber, I. Tsitsilin, F. Haslbeck, C. Schneider, N. Bruckmoser and L. Koch from the Quantum Computing group at the Walther Meissner Institute.
\\ \\
This work was funded by the Munich Quantum Valley (MQV) – Consortium Scalable Hardware and Systems Engineering (SHARE), funded by the Bavarian State Government with funds from the Hightech Agenda Bavaria, the Munich Quantum Valley Quantum Computer Demonstrator - Superconducting Qubits (MUNIQC-SC) 13N16188, funded by the Federal Ministry of Education and Research, Germany, and the Open Superconducting Quantum Computers (OpenSuperQPlus) Project - European Quantum Technology Flagship.

\bibliography{paper.bib}

\end{document}